# Cluster Diffusion and Coalescence on Metal Surfaces: applications of a Self-learning Kinetic Monte-Carlo method

Talat S. Rahman[1*], Abdelkader Kara[1], Altaf Karim[1], Oleg Trushin[2]
[1]Department of Physics, Cardwell Hall, Kansas State University, Manhattan, KS 66506
[2]Institute of Microelectronics and Informatics, Russian Academy of Sciences, Yaroslavl 150007, Russia.
*email:rahman@phys.ksu.edu

## ABSTRACT

The Kinetic Monte Carlo (KMC) method has become an important tool for examination of phenomena like surface diffusion and thin film growth because of its ability to carry out simulations for time scales that are relevant to experiments. But the method generally has limited predictive power because of its reliance on predetermined atomic events and their energetics as input. We present a novel method, within the lattice gas model in which we combine standard KMC with automatic generation of a table of microscopic events, facilitated by a pattern recognition scheme. Each time the system encounters a new configuration, the algorithm initiates a procedure for saddle point search around a given energy minimum. Nontrivial paths are thus selected and the fully characterized transition path is permanently recorded in a database for future usage. The system thus automatically builds up all possible single and multiple atom processes that it needs for a sustained simulation. Application of the method to the examination of the diffusion of 2-dimensional adatom clusters on Cu(111) displays the key role played by specific diffusion processes and also reveals the presence of a number of multiple atom processes, whose importance is found to decrease with increasing cluster size and decreasing surface temperature. Similarly, the rate limiting steps in the coalescence of adatom islands are determined. Results are compared with those from experiments where available and with those from KMC simulations based on a fixed catalogue of diffusion processes.

## I. INTRODUCTION

The topic of this MRS symposium "Modeling of Morphological Evolution on Surfaces and Interfaces," is timely and important because of its relevance to the development of an understanding of microscopic processes that control thin film growth and its temporal evolution. Such theoretical and computational studies nicely complement and supplement experimental observations in the area. Together this three-pronged approach is necessary if we are to build materials whose properties we can control and predict. This is not an easy task because studies of systems of realistic dimensions demand seamless integration of information obtained at the microscopic level into formulations which predict and characterize behavior of systems at the macroscopic scale. We are speaking here of differences of many orders of magnitude. Phenomena at the atomic level extend themselves over nanometers with characteristic time scales lying in the range of femto ($10^{-15}$) to pico ($10^{-12}$) seconds, while thin films for industrial applications are of mesoscopic (~microns) or macroscopic (>millimeter) dimensions and typically take milli-seconds or seconds or even hours to grow and evolve morphologically. Multiscale modeling which has become popular these days remains as yet a challenge, although the field is advancing fast. To date theoretical and computational studies of morphological evolution of surfaces proceed along one of several approaches. There are, for example, macroscopic approaches in which elasticity theory and formalisms of continuum mechanics [1,2] provide an understanding of macroscopic phenomena at solid surfaces. On the other hand, models based on mean field theory and rate equations [3] make more explicit reference to microscopic processes through scaling laws and their comparison with experimental data. For details of some of the achievements of these as well as hybrid models, the reader is referred to a recent review article by Ratsch and Venables [4].

At the other end of the spectrum, fundamental studies of surface morphological evolution are being carried out at the atomistic level using as accurate a technique as feasible. As expected these studies are replete with complex and competing events. Consider, for example, the case of epitaxial growth in which atom adsorption, may be followed by the diffusion of the atom (called adatom) on the terrace, or its nucleation, or the attachment of an adatom to an existing island, or the reverse process of an adatom detachment from an existing island. In the same spirit, the adatom may diffuse along a step edge, or down the step, or nucleate on top of an island. The diffusion of the dimer, trimer, as well as, that of clusters with larger number of atoms or vacancies, may also proceed with significant rates. The nucleation of dimers, trimers, and other adatom and vacancy clusters themselves provide further avenues for anisotropic diffusion since the steps, edges, and corners formed by them may not be symmetric in geometry or in energetics. Stochastic processes like the fluctuations of step edges and dynamical processes which may dominate the relative stability of steps and other defects may offer other avenues for complex growth patterns. Realistic modeling of the evolution of surface morphology has to account for these and other processes as they unveil themselves.

There are thus several key tasks to be undertaken each of which is a challenge in itself. The first of these is accurate determination of the energetics and dynamics of the system at the microscopic level. In this regard we are fortunate to have methods like *ab-initio* electronic structure calculations for the extraction of activation energy barriers and other relevant energetics and dynamics of selected systems of interest. Such calculations are becoming feasible for complex systems, even though they remain computationally intensive. A reasonable alternative albeit not as reliable or accurate, has been provided by several genres of many body interatomic potentials. These potentials have already provided a wealth of information on the microscopic properties of a selected group of metal surfaces which have been tested by comparison with experimental data. With these interatomic potentials it has been possible to carry out computational and theoretical studies of a range of surface phenomena using techniques like molecular dynamics (MD) and kinetic Monte-Carlo (KMC) simulations. While molecular dynamics simulations carried out with reliable interatomic potentials are capable of revealing the essential details of microscopic phenomena as they unfold as a function of temperature, pressure and other global variables, they are limited to time scales (microseconds) which are many orders of magnitude smaller than those for events taking place in the laboratory. For examples, epitaxial growth and surface morphological changes take place in minutes and hours. Furthermore, key atomistic processes that may eventually control the growth pattern and ensuing characteristics of systems are "infrequent" events in the time scales so far accessible to standard molecular dynamics techniques. Recently several attempts [5] have thus been made to overcome this huge difference in time scales by finding ways in which rare events are forced to appear more frequently.

The basic ingredients in atomistic modeling of surface morphology are thus linked with those responsible for the characterization of the diffusion of adatoms, vacancies, and their clusters on surfaces with specific crystallographic orientations and marked with defects and other local environments. When diffusion is driven by thermally activated processes, entities move on a temperature dependent, dynamical surface provided by the substrate. The diffusing entities vibrate about their equilibrium positions and occasionally overcome the energy barrier to move to another site of low occupation energy. To mimic the evolution of surface morphology, we need first and foremost a tabulation of all possible diffusion pathways, and the probability (or rate) with which a particular path (or process) might be undertaken. One way to obtain such information is through molecular dynamics (MD) simulations. But, as straight forward as the method is, it has drastic limitations which leave it uncompetitive for such studies.
An alternative to MD simulations is offered by the kinetic Monte-Carlo (KMC) technique. Kinetic Monte Carlo (KMC) is an extremely efficient method which may be used to carry out dynamical simulations of stochastic and/or thermally activated processes when the relevant activated atomic-scale processes are known [6,7,8]. KMC simulations have been successfully used to model a wide variety of dynamical processes ranging from catalysis to thin-film growth [9,10]. In particular, for problems such as thin-film growth in which the possible rates or probabilities for events can vary by several orders of magnitude, the kinetic Monte Carlo algorithm can be orders of magnitude more efficient than Metropolis Monte Carlo [11]. One of the objectives of this work is to illustrate a new approach to KMC simulation which is expected to provide it with more accuracy and predictive capacity than is presently feasible. The other objective of this paper is to provide some insights into atomistic processes that control surface morphological evolution as found through the application of this technique. Continuing technological developments in experimental techniques like scanning tunneling microscopy (STM) have motivated a good deal of theoretical and computational work to accompany the emerging data on spatial and temporal evolution of small and large atomic and vacancy clusters and their coalescence. Several papers have already addressed some of the issues, particularly with reference to phenomena occurring on fcc(100) surfaces [12]. Some attention has also been paid to questions that arise on fcc(111) surfaces for which STM data [13,14] provoke new thoughts for their analysis. The fcc(111) surface has been enigmatic because of the lack of corrugation in its potential energy landscape which leads to competition between several different types of atomic events and to non-uniqueness in the reaction paths. Naturally, the subject of morphological evolution at surfaces is rich with complex phenomena and vast in its implications for thin film growth. This paper is an attempt to propose a methodology that may provide us predictive power from realistic simulations of such phenomena. After highlighting in section II some of the basic ingredients needed for such atomistic modeling, a summary of the proposed new approach to kinetic Monte Carlo simulations is presented. This is followed in section III with applications of the technique to study small cluster diffusion and coalescence on metal surfaces. Some conclusions and thoughts for future directions are presented in section V.

## II COMPUTATIONAL METHODOLOGY

We present first some details of standard KMC technique and then a summary of the proposed Self-Learning kinetic Monte Carlo Method that we are proposing (SLKMC).

### II.1 KINETIC MONTE CARLO SIMULATIONS

The goal of kinetic Monte Carlo (KMC) is to mimic real experiments through sophisticated simulations. For these simulations to be realistic, it is essential that the system have the flexibility to perform both obvious as well as intricate moves (processes) which may defy common intuition and involve complex scenarios. These processes and their associated energetics and diffusion pathways lie at the heart of a KMC simulation of the time evolution of a given system. To illustrate the point, consider a system containing N particles at a given time with $N_e$ possible types of processes. Let us also associate with each process (i), the number $n_i$ (the number of particles in the system that are candidates for this process). For a given process, the diffusion rate is invariably obtained through the usage of transition state theory [which assumes that the process takes place through a well defined saddle point on the potential energy surface on which the diffusing entity is moving. The diffusion rate for process 'i' is then given by:

$$D_i = D_{0i} \ \exp(-\Delta E_i/k_B T). \tag{1}$$

where $\Delta E_i$ is the activation energy for the process, $k_B$ is Boltzmann constant, T is temperature, and $D_{0i}$ is the so-called pre-exponential or prefactor for the particular process. The total diffusion rate is then given by:

$$R = \sum_{i=1}^{N_e} R_i$$, where, $R_i = n_i D_i$, is the macroscopic rate associated with process i.

In KMC simulations, the acceptance of a chosen process is always set to one. However, the choice of a given process is dictated by the rates. First, a process is chosen according to its probability $P_i = R_i/R$ and next a particle is randomly chosen from the $\{n_i\}$ set to perform this process. This procedure constitutes one KMC step.

### II.2 SELF-LEARNING KINETIC MONTE CARLO (SLKMC) METHOD

From the above description, one can see that the total rate (and hence a full identification) is needed in order to determine the individual probabilities. For complex systems like those associated with epitaxial growth, the number of particles and the number of processes can be very large and simulations may become intractable. To simplify the problem, one assumes that there are only a handful of "*important*" processes that govern the growth and morphological evolution of the system. These types of simulations have been implemented successfully for simple systems but obviously are not suitable for situations in which complex processes involving concerted motion of several atoms in a wide variety of environments are present. A rethinking of the way we perform KMC has become a necessity. Simulations with an *a priori* chosen catalogue of processes need to be replaced by a continuous identification of possible processes as the environment changes.

However it has been shown theoretically that many-particle processes can play important role in explaining mass transport on Cu(100) [15,16] and thus should not be ignored. There are also experimental evidences that many particle processes should take place on FCC(111) surface too [17,18]. In complex environments intricate single atom processes previously not encountered may also become relevant. Recently several efforts have been made to overcome some of these deficiencies. Some acceleration schemes [5] that have so far been proposed include parallel replica dynamics [19], hyperdynamics [20], temperature accelerated dynamics (TAD) [21], and on-the-fly KMC [22,23]. While the goals are similar, the approaches are different and their feasibility depends on the type of approximations that have to be made. For example, one point of departure in the proposed techniques is whether transition state theory (and the harmonic approximation) is imposed and whether pre-exponential factors are calculated. Another point of difference is whether off-lattice events are allowed, for example fcc to hcp occupancy. Another is wether the lattice itself is necessary. These methods and their combinations have already provided valuable information about matters such as the importance of multi-atom events and the differences in prefactors for critical events in thin film growth [22]. In particular the work of Henkelman and Jonsson [22], confirms the significant role played by multi-atom or collective processes which may also be accompanied by unusual prefactors (entropic term). Their work also indicates the importance of large sampling of the phase space of the system to capture key atomistic processes. We find the approach of Henkelman and Jonsson to be useful but of limited application because of the need to perform a large number of trial trajectories at every KMC step and without the opportunity to use this information again, at a later point in the simulation. We believe that by applying a pattern recognition scheme, we will be able to train our model systems to learn from previous diffusion (self-learning) paths and trajectories that we store in a data base.

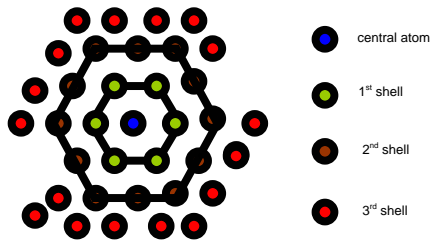
Fig.1: Illustration of the 3-shell scheme for an (111) fcc 2D system.

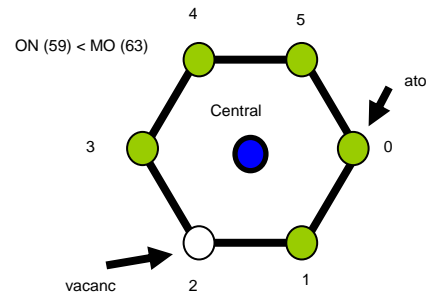
Occupancy Number (ON) = $1 \times 2^0 + 1 \times 2^1 + 0 \times 2^2 + 1 \times 2^3 + 1 \times 2^4 + 1 \times 2^5$

ON (59) < MO (63)

Fig.2: The central atom is considered 'active' if the occupancy number (ON)

It is obvious that in this new way of performing KMC simulations, the local environment is the key issue and should be the base ingredient. In order to illustrate the essential of SLKMC simulations of complex systems, we choose the fcc(111) surface with a six-fold symmetry as an example. We assume that any process in this system involves a central atom and, in the case of concerted motion, atoms in the next 3-shells as illustrated in Fig. 1. We further assume, without loss of generality, that a process may be described in terms of the central atom moving to a neighboring vacancy while allowing all atoms in the 3 surrounding shells to participate in the process. Thus multi-atom processes will be allowed to appear at par with those involving only single atoms. An important book-keeping aspect lies in the labeling of the three shells which is done first in binary and subsequently a base ten number is associated with each shell. Hence, for an atom in the system to be active (i.e. the central atom for a given process), it should have a vacancy in its first shell (or an occupancy number less than 63) as illustrated in Fig.2.

Once the procedure for classifying atoms as active and non-active is completed for all the atoms in the system, we proceed by determining all the possible processes associated with every active atom. This step is then followed by the determination of the activation energy and pre-factor for all processes. *This is the bottleneck for the simulation*. Even when we make the widely-used assumption that all the processes have the same pre-factor, the calculation of the activation energy is very expensive if one needs accurate values. Note that since the activation energy is in the exponential, any small variation in the activation energy results in a substantial change in the relative probabilities and hence the outcome of the whole simulation. To overcome this bottleneck, we have introduced a "*self-learning*" KMC in which all calculated activation energies are stored in a database using the labeling described above. At the start of every Monte Carlo step, the labeling scheme identifies the initial configurations associated with all possible processes that the system may undergo, and verifies whether the corresponding activation energy barriers exist in the database. If a new configuration is encountered (due to morphological changes), the associated processes and their activation energies are computed and stored in the database. A complete list of microscopic ($D_i$) and macroscopic ($R_i$) rates is then tabulated and the Monte Carlo step is completed.

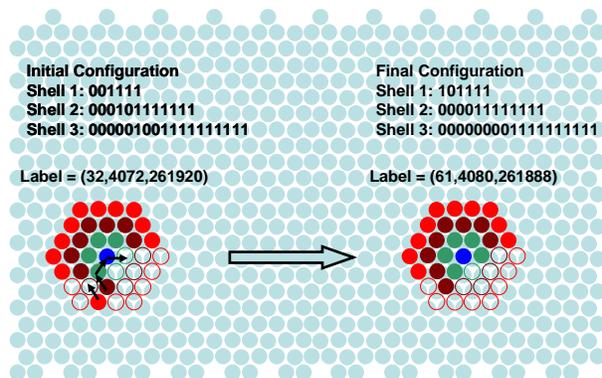

Fig. 3 Schematic representation of the re-labeling procedure in the 3-shell scheme, during multiple atom processes.

As an example of multi-atom processes that appear in the database, we show in Fig.3 the concerted motion of four atoms. Note that this is counted as one single step process identified by the labeling procedure described above. Since the role of multi-atom processes has been the subject of much discussion, we present in Fig. 4, their relative importance as a function of cluster size and temperature for Cu adatom island diffusion on Cu(111) discussed in section III.1.

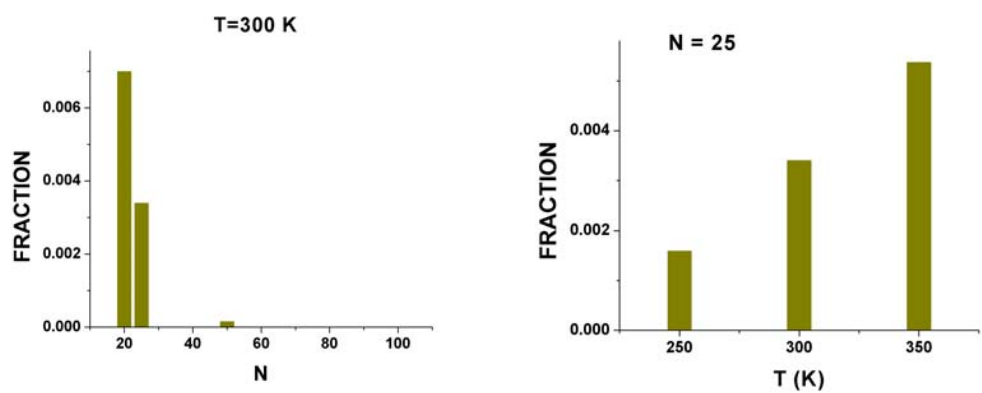

**Fig 4.** Frequency of multi-atom processes as function of size at 300K (left), and of temperature for a 25 atom-cluster (right).

## III. APPLICATION OF SLKMC TO CLUSTER DIFFUSION AND COALESCENCE

Experimental studies of the diffusion of adatoms and small atomic clusters on metal surfaces using Field Ion Microscopy (FIM) have already provided a number of unexpected events such as the concerted motion of atoms [24] and the collective sliding motion of clusters [25]. Moreover, STM measurements on Ag(100) [26] have established that large adatoms move and those on Ag(111) [13] have confirmed that the mobility of the vacancy islands is comparable to that of the adatom islands. In detailed studies of post-deposition surface morphological evolution Giesen and co-workers [14] have also made interesting observations on the coalescence of two-dimensional Cu islands on Cu(111). These and related observations led to a series of papers [27-30] with speculations about the microscopic mechanisms that cause these islands to move. Of particular interest here are the competing mechanisms of adatom periphery diffusion, evaporation/condensation, and terrace diffusion. Statistical mechanical calculations based on solid-on-solid (SOS) models predict specific scaling of the diffusion coefficient with the island diameter, depending on the preponderance of one of these three mechanisms. Since these dependencies are not unequivocally extracted from experimental data, because of the large error bars involved, the issue is not yet completely settled, although the bias is towards periphery diffusion. Molecular dynamics simulations of Ag vacancy island on Ag(111) [31] have shown a preference for periphery diffusion but the conclusion cannot be definitive because of the limitations of MD, as already discussed. Questions about the validity of elastic-continuum based models, led Bogicevic *et al* [32] to carry out KMC simulations of these systems using a small number of diffusion processes. They find that the exponents in the power law dependence of the diffusion coefficient on island size were themselves temperature dependent and material specific, unlike earlier predictions. While the work of Bogicevic *et al* points to the simplicity calculations preceding theirs, it also begs the question whether atomistic models based on a few hand-picked diffusion processes are capable of displaying the inherent complexity of the system. The issue is whether the evolution of the system could be prejudiced by the usage of an insufficient set of atomic processes arising from a narrow local consideration. Below we present some results of our simulations. The point of departure in our work is the usage of pattern recognition schemes in SLKMC simulations which allows the creation of a data base containing the events and their energetics that the system requires for unbiased evolution, as discussed in the preceding section.

### III.1 Some results from small two dimensional cluster diffusion on Cu(111)

Below we present some results of simulations of the diffusion of 2D Cu islands on Cu(111), containing 10-100 atoms, for about 500 million MC-steps at several temperatures. These simulations were performed with the open data base (SLKMC) until the system evolution reached equilibrium conditions, as judged by a count of nearest neighbor bonds of the active atoms and the behavior of the mean square displacement of the cluster center of mass. The results are compared with those from a standard KMC simulation in which a total of 294 (49x6) processes involving single-atom peripheral diffusion were utilized [33,34]. Since the physical time elapsed at each MC-step is governed by the rate of the process, they are unequal in length. Thus, to calculate the mean square displacement of the center of mass, we filter our data to a set of almost equidistant MC-step along with the corresponding center of mass coordinates.

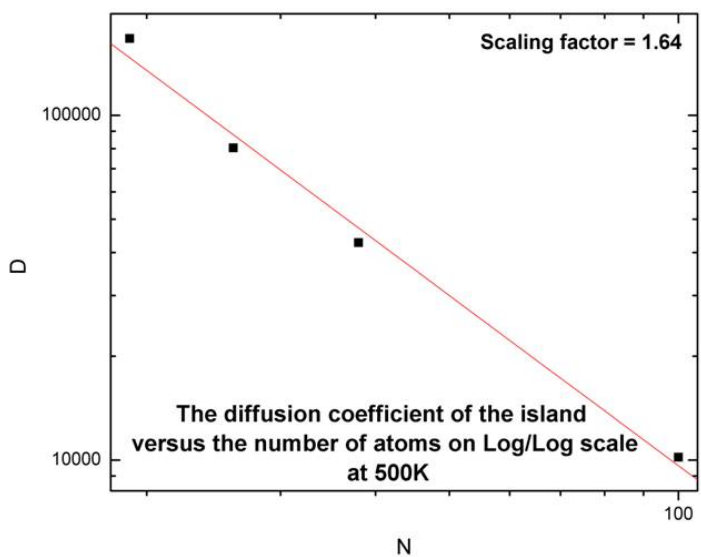

**Fig. 5.** The variation of the diffusion coefficient D with cluster size at 500 K (log-log plot).

The plot in Fig. 5 shows that for the four sizes considered SLKMC simulations predict a scaling factor of 1.64 for the diffusion coefficient at 500 K ($D \sim N^{-1.64}$). At 300 K the scaling factor is 1.57. The variation of the scaling factor with temperature thus lies within the statistical error bars in our simulations and for all intents and purposes it may be assumed to be independent of temperature. A simple look at this result would imply that these clusters move because of single atom periphery diffusion and not much new has resulted from the usage of SLKMC. To get a better grasp of the details of the atomistic events involved and any differences that may have come about because of the freedom that the system had in the choice of diffusion mechanisms including those involving multiple atoms, we present in Table I the results obtained above with those from standard KMC simulation with a predetermined catalogue of processes.

**Table I**. Diffusion Coefficients for clusters of four sizes at 300 K and 500K from SLKMC and standard KMC

| Cluster size (atoms) | Diffusion coefficient D ($A^2$/sec) | | | |
|---|---|---|---|---|
| | 300K | | 500K | |
| | SLKMC | KMC | SLKMC | KMC |
| 19 | 0.196 | ---- | $1.67 \times 10^5$ | $1.29 \times 10^4$ |
| 26 | 0.170 | 0.48 | $8.05 \times 10^4$ | $2.24 \times 10^4$ |
| 38 | 0.117 | ---- | $4.27 \times 10^4$ | $2.22 \times 10^4$ |
| 100 | 0.016 | ---- | $1.02 \times 10^4$ | $3.20 \times 10^3$ |

Quite clearly Table I shows remarkable differences in the diffusion coefficients obtained by the two types of simulations. In the case of the 19 atom cluster the diffusion coefficients differ by an order of magnitude: single atom periphery diffusion underestimates the mobility of this perfect hexagon. For the other sizes the differences are noticeable. Some insights for these differences can be obtained from Table II below in which we have summarized the frequencies of the processes that are executed in the two types of simulations. While the most dominant mechanism is that of a single adatom along the A-type ((100)-microfacetted) step edge in both SLKMC and KMC, there are differences not only in the frequencies which the various processes are executed in the two types of simulations, but also the appearance of several new processes like detachment from the corners engulfed by the A and B-type ((111)-microfacetted) step edges, as shown below in the figures with the appropriate activation energy barriers calculated using the drag method. In Table II the activation energy barriers obtained from the drag method are compared with those from the nudged elastic band (NEB) method. Note that in all cases we have used interatomic potentials from the embedded atom method [35]. Further details of the processes listed in Table II, including their nomenclature, can be found at (http://www.phys.ksu.edu/~rahman). Note that even though multiple atom processes are not very frequent, they do occur, as shown erlier in Fig.4, and in some cases they may be the rate limiting step for the cluster diffusion.

**Table II:** Frequency of Processes for the 19 atom cluster (Hexagon) at different temperatures

| Temperature  Processes | Energy Barrier (eV) NEB | Energy Barrier (eV) Drag | 300K KMC | 300K STKMC | 500K KMC | 500K STKMC |
|---|---|---|---|---|---|---|
| Step Edge A | 0.252 | 0.250 | 0.62 | 0.6797 | 0.42 | 0.511 |
| Step Edge B | 0.295 | 0.310 | 0.17 | 0.0954 | 0.24 | 0.1403 |
| Kink Detach along Step A | 0.519 | 0.521 | 0.0 | 0.0 | 0.0020 | 0.0016 |
| Kink Detach along Step B | 0.556 | 0.538 | 0.0 | 0.0 | 0.0 | 0.0008 |
| Kink Detach along Step (small) A | 0.608 | 0.620 | 0.026 | 0.0106 | 0.012 | 0.0 |
| Kink Detach along Step (small) B | 0.680 | 0.693 | 0.0016 | 0.0007 | 0.0023 | 0.0018 |
| Kink Incorp. A | 0.220 | 0.220 | 0.0 | 0.0001 | 0.0020 | 0.0025 |
| Kink Incorp. B | 0.265 | 0.287 | 0.0 | 0.0 | 0.0 | 0.0009 |
| Kink Incorp. (small) A | 0.0075 | 0.009 | 0.025 | 0.0 | 0.011 | 0.002 |
| Kink Incorp. (small) B | 0.0810 | 0.108 | 0.0 | 0.0 | 0.0012 | 0.0 |
| AA corner detachment | **** | 0.440 | **** | 0.0007 | **** | 0.0063 |
| Kink Detach out of Step B | 0.590 | 0.600 | 0.0 | 0.0091 | 0.0 | 0.0098 |
| Kink Fall into Step A | 0.074 | 0.102 | 0.0 | 0.0007 | 0.0 | 0.0016 |
| Kink Fall into Step B | 0.0069 | 0.015 | 0.0 | 0.0109 | 0.0 | 0.0101 |
| BB corner detachment | **** | 0.344 | **** | 0.0322 | **** | 0.0451 |
| All multiple atom processes |  |  | **** | 0.00015 | **** | 0.0042 |
| KESE A | 0.374 | **** | 0.0 | **** | 0.0011 | **** |
| Corner Rounding at AA stage 1 | 0.313 | 0.325 | 0.0 | 0.0001 | 0.0 | 0.0017 |
| Corner Rounding at AA stage 3 | 0.0096 | 0.014 | 0.0 | 0.0 | 0.0 | 0.0017 |
| Corner Rounding at BB stage 1 | 0.374 | 0.393 | 0.0 | 0.0 | 0.0 | 0.0002 |
| Corner Rounding at BB stage 3 | 0.052 | 0.072 | 0.0 | 0.0 | 0.0 | 0.0002 |
| Corner Rounding at AB stage 1 | 0.317 | 0.328 | 0.066 | 0.0579 | 0.11 | 0.0894 |
| Corner Rounding at AB stage 2 | 0.0839 | 0.113 | 0.0053 | 0.0023 | 0.024 | 0.0158 |
| Corner Rounding at BA stage 1 | 0.396 | 0.421 | 0.0047 | 0.0013 | 0.023 | 0.0095 |
| Corner Rounding at BA stage 2 | 0.0148 | 0.021 | 0.067 | 0.0884 | 0.12 | 0.1348 |
| AB corner detachment towards B step | **** | 0.619 | **** | 0.0003 | **** | 0.0017 |
| AB corner detachment towards A step | **** | 0.689 | **** | 0.0 | **** | 0.0002 |

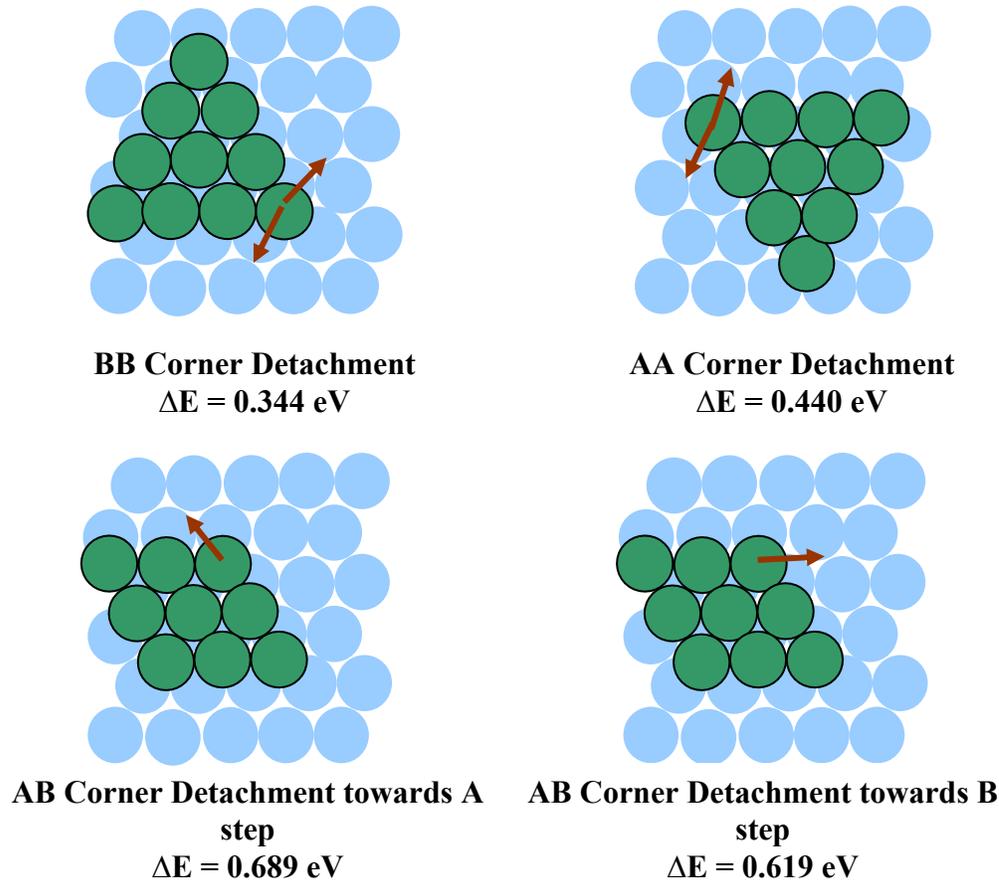

**BB Corner Detachment**
ΔE = 0.344 eV

**AA Corner Detachment**
ΔE = 0.440 eV

**AB Corner Detachment towards A step**
ΔE = 0.689 eV

**AB Corner Detachment towards B step**
ΔE = 0.619 eV

**Fig. 6** Key processes and their corresponding activation energy barriers, from EAM potentials.

### III.2 Some results from island coalescence

As a second example of application of SLKMC simulation with the open data-base, we present here results of the coalescence process in which two adatom islands join together to form a larger island with an equilibrium shape on Cu(111). Successive snapshots of the system during KMC simulation are shown in Fig 7. The islands consisted of 78 and 498 atoms in almost hexagonal forms which after several million KMC steps start sharing a neck, then form an elongated island, and finally settle into triangular shape which ultimately settles into a hexagon. This is a remarkable result as our simulations show almost perfect agreement with the experimental observations of Giesen et al.[33]. Note that in these simulations the system was free to evolve with the diffusion mechanisms of its choice.

## IV. CONCLUSIONS

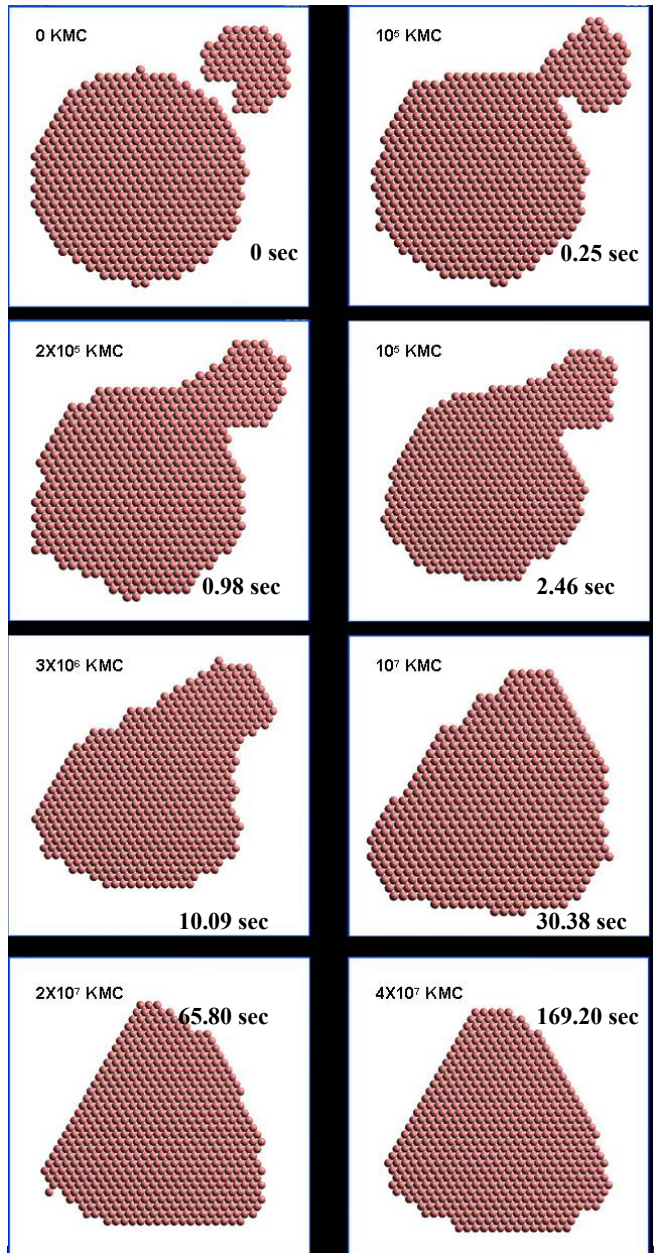

**Fig. 7** Cluster coalescence at 300K, small cluster contains 78 atoms and big cluster contains 498 atoms.

The above sections provide a brief summary of some of the techniques that are used in atomistic modeling of thin film growth and its morphological evolution. The field is still in its infancy as accurate methods like *ab initio* electronic structure calculations are only now becoming feasible for systems with as much complexity as those presented here. Once activation energy barriers of all relevant processes and their diffusion paths can be obtained from such methods, KMC simulations appear to provide an attractive procedure for predicting and understanding the characteristics of thin films as a function of their atomistic structure, substrate crystallography, and temperature. As we have already alluded to, the task of calculating diffusion prefactors is still ahead of us. This is particularly important since we find many competing processes to differ only slightly in energy and differences in their vibrational entropy contributions to the prefactors can make a difference in the ultimate evolution of the film morphology. Another important result from our simulations with the open data base is that dynamical evolution of the system with prejudged diffusion processes may yield erroneous results. Also, the pattern recognition schemes to be a prudent way to develop data base of diffusion processes and their energetics. It does involve a lot of work in the beginning but once the data base is compiled, it can be used for any type of simulation of the system. Of course, for realistic simulations of thin films we need to incorporate exchange and other processes which involve motion in 3D. Such effort is currently underway.

**ACKNOWLEDGEMENTS**
This work was supported partially by a grant from NSF (ERC-0085604) and from CRDF (RU-P1-2600-YA-04). We thank James Evans, Ted Einstein, and Ahlam Al-Rawi for many helpful discussions.